# Speaking Plant Approach for Automatic Fertigation System in Greenhouse

Usman Ahmad*[1], Dewa Made Subrata[1], and Chusnul Arif[2]

[1]Dept. of Mechanical and Biosystem Engineering – Bogor Agricultural University (IPB), Darmaga Campus, Bogor 16680
usmanahmad@ipb.ac.id, dewamadesubrata@yahoo.com

[2]Dept. of Civil and Env. Engineering – Bogor Agricultural University (IPB), Darmaga Campus, Bogor 16680
chusnul_ar@yahoo.com

*Abstract*

*Nowadays, many vegetables are grown inside greenhouses in which environment is controlled and nutrition can be supplied through water supply using electrical pump, namely fertigation. Dosage of nutrition in water for many vegetable plants are also known so that by controlling water supply all the needs for the plants to grow are available. Furthermore, water supply can be controlled using electrical pump which is activated according to the plants conditionin relation with water supply.*

*In order to supply water and nutrition in the right amount and time, plants condition can be observed using a CCD camera attached to image processing facilities to develop a speaking plant approach. In this study, plants development during their growing period are observed using image processing. Three populations of tomato plants, with less, enough, and exceeded nutrition in water, are captured using a CCD camera every three days, and the images were analyzed using a developed computer program for the height of plants. The results showed that the development of the plants can be monitored using this method. After that, the response of plant growth in the same condition was monitored, and the response was used as input for the fertigation system to turn electrical pump automatically on and off, so the fertigation system could maintain the growth of the plants.*

## 1. Introduction

Vegetables are very important as source of vitamins and fibers. To increase vegetables production, technology in vegetables cultivation is continually improved. Such a technology is growing vegetables in relatively controlled space in a greenhouse or screenhouse. Furthermore, some materials for constructing greenhouse are modified so the greenhouse can be optimized to trap only a certain of wavelength from sunlight for specific purposes [1]. Nowadays, in Indonesia many types of vegetables are grown in a well controlled environment such as in greenhouse and screenhouse. Usually, the vegetables are grown on a normal sterilized medium like soil and paddy husks, charcoal, sand, small rock, carbon, or zeolit, as long as they are sterilized [2].

Some experiments for determining dosage of nutrition needed for specific plants were also conducted and the results have been reported . For strawberry, the suggested dosage are 30 lb/a of nitrogen, until 100 lb/a of phosphate depends on the soil condition, until 100 lb/a of potassium depends on the soil condition, 15 lb/a of sulphate, 20 lb/a of magnesium, and small amount of boron [3]. It was also reported that mixing of nitrogen, phosphate, potassium, magnesium and zing then applied them to potato gave a better result [4]. For red chili, recommended dosage for optimum result is 100 kg/ha of nitrogen, 90 kg/ha of phosphate, 90 kg/ha of potassium, 20 kg/ha of sulphate, and 2 kg/ha of zing [5]. For tomato, recommended dosage for optimum result is 175 kg/ha of urea, 350 kg/ha of TSP, and 200 kg/ha of KCl [6]. Other work reported that nutrition needed to grow tomato is 8-10 g/plant for urea and 10-15 g/plant for TSP [7].

Plant nutrient for horticulture commercially available both in solid and liquid forms. Nowadays, liquid form is preferable because of simple and easy to apply, especially for vegetables, ornamental, and some kinds of fruits. It is believed that more than 100 different brands are commercially available in Indonesia with primary and secondary nutrition as well as micro compounds [8]. Every brand has its own unique formula for several plants which is determined by nutrition composition and concentration.

Supply of plant nutrition in water can be done continuously by flowing the water into the medium where the plants are growing, but this method will consume a lot of electrical power to run the pumps all the time. Besides, some amount of plant nutrition in drained water will be released to the surrounding and this increase the production cost and affect the environment. One method to avoid this from happening is by giving the nutrition in water only when the plants need it, not all the time. And the plants might need different amount of plant nutrition and water from day to day during the growing period since they might experience different micro climate in the greenhouse as a factor of external climate changes. We might be able to predict when and how much plant nutrition and water is needed by the plants if we know response of the plants to micro climate changes, or normal growth rate of the plants during growing period. One possibility to observe normal growth rate of the plant is by taking its images during its growing period and take some parameter as a series of data using image processing technique. Also, to know response to the micro climate changes, the plant can be placed in different conditions and take its images, analyze them and compare the data to the optimal condition.

Whatever the medium is, this method of cultivation needs plant nutrition that can be supplied through watering by using electrical pumps, namely fertigation system. Hence, plant nutrition supply through fertigation system for vegetables grown in the greenhouse is very important to obtain good produces at harvest time. For this purposes, dosage of nutrition for some vegetable plants are known, and the formula is already developed for each plant. What we have to do is to put some amount of plant nutrition into a gallon of water for example, stir and distribute it to the plants. If the nutrition in water is given in a right amount and time during the growing period, the plants will grow and produce whatever we want from them in a normal period of time, with a good yield and quality. To accomplish this task, application of automatic control would be very helpful, which is sometimes referred as biosystems, meaning that automatic application using computer technology and electronic in agricultural activities [9].

Automation in agricultural activities is considered necessary based on the following reasons; 1) many tasks are still laborious and monotonous which are not suitable for human, but require certain intelligence to perform, 2) the availability of farming workforce is decreasing at an alarming rate in many countries, 3) the problem of labor shortage frequently result in rising in labor cost, and 4) the market demand for product quality has became an important factor in bioproduction [10]. Image processing is a technology to get information from images by manipulating images and produces the desired information to be used for taking action, which is very useful in process automation. When image processing is employed and collected information is used to operate a device such as electrical pump, it is called machine vision [11].

Conventional greenhouse production in agriculture relies on environmental sensors and often sophisticated algorithms to control climate and irrigation. Sensors for incident light, temperature, and humidity, provide climate information while sensors for pH, electrical conductivity (EC) and media moisture content provide information on the root-zone environment. These sensors furnish information for record keeping and control purposes. However, they provide only indirect information on the status of the crop. Physiological information could provide important supplementary data, especially if integrated into control systems or computer models in the so-called speaking plant approach [12][13]. Development of control system based on the feedback from the plants is called plant speaking system. To develop a plant speaking system, growth characteristics of the plant and its relationship with nutrition and water should be studied first [14]. For example, growth rate of groups of lettuces grown in different light condition were captured and analyzed to determine the optimum environment [15].

The concept of speaking plant approach (SPA) was proposed and the SPA-based intelligent control technique consisting of a decision system and feed back control system was applied to the optimizations of hydroponic tomato cultivation. The decision system consisting of neural network and genetic algorithms allowed the optimal values (optimal number of steps for set points of environmental factors) that maximize (or minimize) plant responses to be successfully obtained. In the hydroponic tomato cultivation, the optimal 4-step drainage and supply times for hydroponic solution were determined to be 4 min drainage, 8 min supply, 4 min drainage, 2 min supply. This operation apparently increased the net photosynthetic rate of a tomato. On the other hand, the optimal 4-step set points of the nutrient concentration during seedling stage were decided to be 1.4, 0.3, 1.6, 2.0 mS/cm. The total leaf length to stem diameter ratio (TTL/SD) was 10-15% higher with this optimal operation than with a conventional method. The results suggest that a control method changing flexibility and optimally on the basis of plant responses in a better way to improve the quality of the plants during cultivation than a conventional control manner keeping constant at the adequate level, and SPA-based intelligent control techniques are suitable for the dynamic optimization of a total plant production system [16].

Image processing has been developed and tested in many agricultural activities; for example in automatic cherry tomato harvester in greenhouse [17], mushroom picker to find and locate mushroom when it is ready to harvest [18], and watermelon harvester to find watermelon and judge it for harvest criteria [19]. Color images processing was used to predict ripeness level of Gedong mango in a sorting and grading machine [20]. Also, an image-processing algorithm was developed for a ground-based remote sensing system to determine the position of the leading edge of water in an irrigation furrow without entering the field. The horizontal plane ranging method, using the number of pixels between furrows at the leading edge of water in an image, was successful in determining the distance of the leading edge of water from the end of the field with an average error of 1.2 m and a 95% confidence limit of ±2.72 m. The method developed had the following characteristics; hue segmentation was a robust means of detecting water in the furrow for a range of furrow orientations and sun angles under sunny and partly cloudy conditions, for fully overcast skies, intensity segmentation successfully detected water in the furrows when hue segmentation failed, metric rectification was used to obtain accurate estimates of the pixel distance between furrows before the crop emerged. The method was robust for the amount of water in the furrows, allowing accurate extrapolation of the furrow centerline even when a limited amount of water was visible in the furrow, once the crop emerged, hue segmentation detected the crop plants allowing pixel distances between crop row centerlines to be determined without metric rectification, and errors in determining the distance of the leading edge of water from the end of the field were not significantly different between metric rectification and crop plant segmentation methods [21].

In order to obtain data from plants during the growth periode, a complete system called CropAssist was described, which may be used to continuously monitor a greenhouse vine crop for water use and 24 hours growth. The system used pairs of load cells and a through system to capture leachate, crop growth and water use, and many irrigation parameters may be measured simultaneously. The equipment is non-obtrusive in both research and commercial greenhouse environments. The system concurrently monitors four sites, with up to 12 plants each, but is expandable as needed with more sites strategically placed in the greenhouse. It may be possible to construct a more elaborate plant support system capable of monitoring an entire row. Based on the CropAssist model, other crop monitoring systems could be developed which bypass the need for a data logger or use other software environments. The data may be easily exported to spreadsheet programs for further analysis or presentation. The CropAssist system offers significant research opportunities in modeling whole-plant behaviour in response to climatic and cultural conditions. The system may also aid in crop management decisions. For example, based on the near real-time (every 5 minutes) display of transpiration rate, greenhouse temperature or relative humidity could be modified to achieve a desired transpiration rate for the crop. Applications to other vertical or vine crops, perhaps even outdoor raspberries and grapes, should be investigated [22].

A video-processing algorithm using maize plant region features in video frame sequences from a commercially available digital camcorder was effective in detecting early growth stage maize plants for population sensing. The research showed that interplant distances measured in pixels from video frames can be used to effectively estimate interplant distance. With statistical separation of weed and maize plants, the overall maize plant count root mean squared error (RMSE) was 2.1 plants or 8.7% in 6.1m row sections across the range of conditions. Plant count estimation error was dependent on tillage treatment, and error variance increased with increasing growth stage. Variance in location estimation error increased as growth stage increased. No evidence of significant differences was found between mean measured and estimated interplant distances for all treatment combinations [23].

The development of a novel smart sensor that can estimate plant-transpiration dynamic variables such as transpiration, stomatal conductance, leaf-air temperature difference, and water vapor deficit in real time was conducted. This smart sensor fuses five primary sensors: two temperature sensors, two relative humidity sensors and one light sensor. To show the effectiveness of the proposed smart sensor, it was compared with a commercial Phytech PTM-48M transpiration monitoring system. The results show that the proposed sensor can obtain very similar results compared to the reference system with less noise due to the digital filtering applied to the primary measurements. The transpiration dynamics variables are calculated in real-time from the primary sensor data providing very useful information related to the plant transpiration which is valuable to schedule irrigation, prevent diseases, and detect drought conditions in precision agriculture. Similar behavior of the estimated transpiration variables shows the relationship between these and how they depend on the primary sensor readings. The necessary computations in order to obtain the transpiration dynamics are computed in a low-cost FPGA platform in which parallel architecture is utilized to implement the transpiration equations. This permits the integration of the data communication, memory management data acquisition and signal processing in a single embedded sensor which can be used to monitor plant transpiration variables and their relationships in a wide range of precision agriculture applications. Finally, transpiration related stress conditions can be detected in real time because of the online processing and communications capabilities. All of which constitutes a very useful precision agriculture smart sensor [24].

Other research showed that a strong non-linearity exists in the relationship between water content of the plant and the value of the finite element feature. Inspire of that, the result indicated that the neural network model proposed was able to correlate two naturally different quantities. It is also clear that the finite element features can be use effectively to quantify a variation of pictorial image [25]. Modelling a biological system by means of mathematical identification techniques can be considered to be more objective than by physical and physiological laws where a system is considered almost forced to behave according to a predefined schedule. That is one of the reasons why black box models could achieve higher accuracy. It should be much easier to assess accuracy and to validate the model. For some of these techniques validation is even incorporated into the modelling process [26].

Basic mechatronic and machine vision principles were applied to develop a variable rate herbicide applicator to optimize the herbicide application rate corresponding to the amount of weeds. The study mainly focused at applying the systematic concepts and by using the low-cost devices locally available in Thailand. The tractor mounted real-time variable rate herbicide applicator performed satisfactorily under actual field conditions. The developed system has the adjustment flexibility for RGB ratio, threshold level of percent green color, threshold level of percent duty of PWM, capture picture size, with forward speed compensation and could be reset by the user before start of the work. Testing of the developed sprayer showed at least 20% reduction in the herbicide amount [27].

The objective of this preliminary research is to observe the growth rate of plants population growing in the greenhouse in different fertigation formula, using image processing technique. Another objective is to develop a real-time monitoring system using a CCD camera that can be used for automatic fertigation system.

## 2. Material and Methods

The research was conducted in experimental field of Mechanical and Biosystem Engineering Department, Faculty of Agricultural Engineering and Technology, Bogor Agricultural University (IPB), Bogor, Indonesia. The research was started on June 2008, and data (images) collecting was accomplished on August 2008. After that, the images were analyzed using a developed computer program. The application of automatic fertigation system was conducted in July 2009.

Material used in this research were tomato seeds, paddy husk charcoal as medium to grow the plants, polybag, and commercially available plant powder nutrition for horticultures. Equipments used were a greenhouse, a set of drip irrigation system with a water tank and elastic pipes. For still images analysis, a digital camera was used for image acquisition, which was taken every three days during the growing period, and a computer with image processing program to analyze the images. For real-time analysis in automatic fertigation system, a CCD firewire camera attached to a computer with image processing program was used.

First, the tomato seeds were grown in a tray in shade to produce seedlings. After two weeks, the seedlings were placed into polybags filled with charcoal of burned paddy husk. Before that, the charcoal was sterilized by soaking it into water with disinfectant. The polybags with seedlings then were placed in the greenhouse and their images were taken individually every three days. There were three groups of plants grown in the greenhouse, every group consist of 20 plants. Different formula in concentration was applied to the three groups, one group with under-fertilized (1.0–1.5 mS or mili Siemens, a unit for concentration measurement through the electrical conductivity), one with normal-fertilized (2.5–5.0 mS), and one with over-fertilized (10.0–12.5 mS).

Since they were placed in the greenhouse, their images were taken every three days from the same distance using a digital camera, with a red panel placed as a background. When the plant grew bigger and taller so that camera could not cover the whole plant from the same distance, the distance was adjusted. Later, in image processing, distance factor was considered so that the data resulted from image processing can be compared from day to day.

In the application of automatic fertigation system, the response of the plant for deficiency in fertigation was also studied from the images captured real-time by the firewire CCD camera. For this purpose, width of the plant in image captured at time **t** is compared with width of the plant in image captured at time **t**+Δ**t** because plant tends to wilt when experiencing water deficiency. Therefore, it is important to determine water deficiency condition by finding the right ratio value of width comparison. Ratio of plant's width for current condition and plant's width of normal condition might be useful for wilt determination using image processing.

## 3. Results and Discussion

A number of tomato seedlings were grown in polybags and placed in the sterilized greenhouse after two weeks of nursery. Image of every plant was captured using a digital camera (Fig. 1). The distance in capturing the plant image was increasing as the plants grew taller and taller (from 30 to170 cm, with 10 cm increment for every 3 days). The captured images of tomato plants during their growing period were then analyzed using a developed image processing program. Prior to analysis, the images were collected and stored in a harddisk, separated into their groups respectively. Every image was analyzed and height of every plant in the images were calculated.

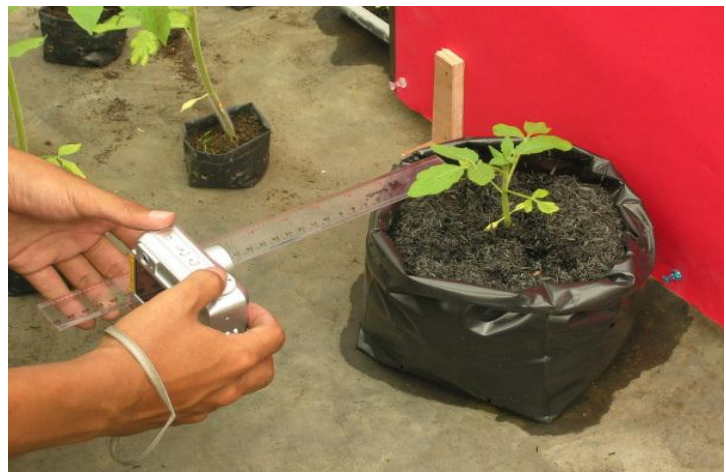

**Figure 1. Image acquisition of tomato plant**

Data obtained from image processing were processed to get average of height for every group (different days of capturing and different treatments for nutrient concentration). The average height data then were plotted into a graph to see the growth rate of the plants during the growth periode (Fig. 2).

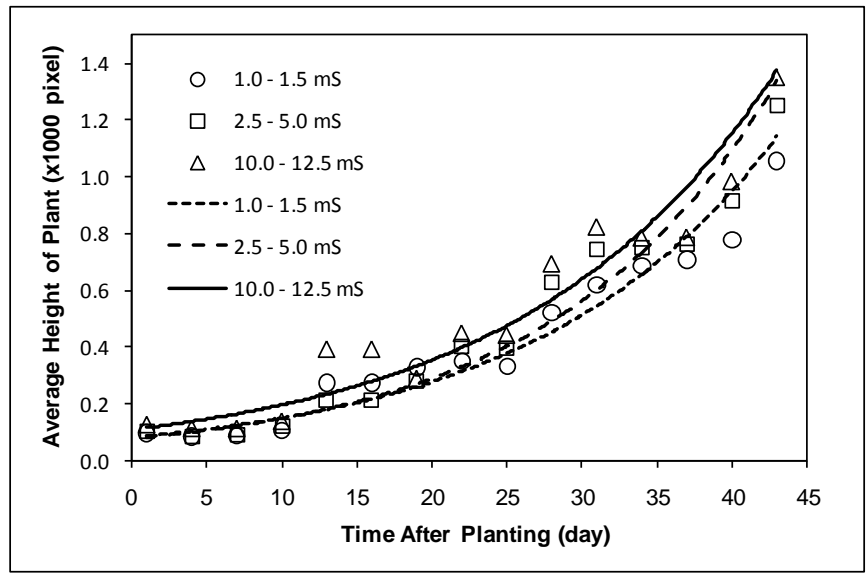

**Figure 2. Plant height development obtained from image processing**

The plants were actually grown until they produced fruits, around 75 days after planting. However, due to small space in the greenhouse, after 43 days the plant could not be captured individually since there were leaves overlapping among the plants. Overlapped leaves in image were difficult to analyzed because the developed program was designed to analyze tomato plant image as it is taken individually.

From Fig. 2, we can see that group of tomato plant with under-fertilized (1.0–1.5 mS), experienced the slowest growth rate compare to the other two groups, the plants with normal-fertilized (2.5–5.0 mS) and the plants with over-fertilized (10.0–12.5 mS). Group of plants with over-fertilized showed the fastest height increment, followed by group of plants with normal-fertilized and group of plants with under-fertilized. Change in image quality due to change of sunlight intensity during image acquisition time, caused some fluctuation data, indicated by decreased values at some points (normally, the height of plants increases from day to day). We can see also that the graph is non-linear, meaning that the tomato plants growth was increasingly faster from day to day. More attention should be taken especially when the plants were very small for the first two weeks after planting. Lack of nutrition in this period will cause abnormal growth where the plants are still tiny and weak in roots system.

Automatic fertigation was attempted by applying real-time monitoring using a CCD camera and plant wilt determination was developed in the computer program to turn pump ON and OFF (Fig. 3). The plant that observed by the camera was placed in front of a red cloth to get contrast background with the plant in image. The camera connected to a laptop with interfacing facility to control an electrical pump, that will pump the fertilized water into all the plants. When the result of image processing meets the criteria of wilt condition, the program will turn ON the pump for 3 minutes, then OFF again.

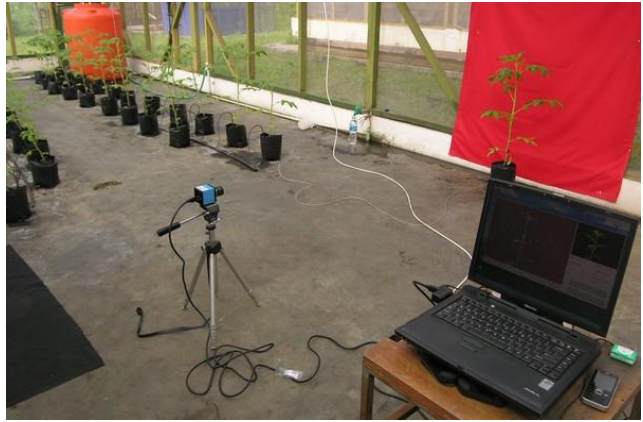

**Figure 3. Real-time plant monitoring system setup in the experiment**

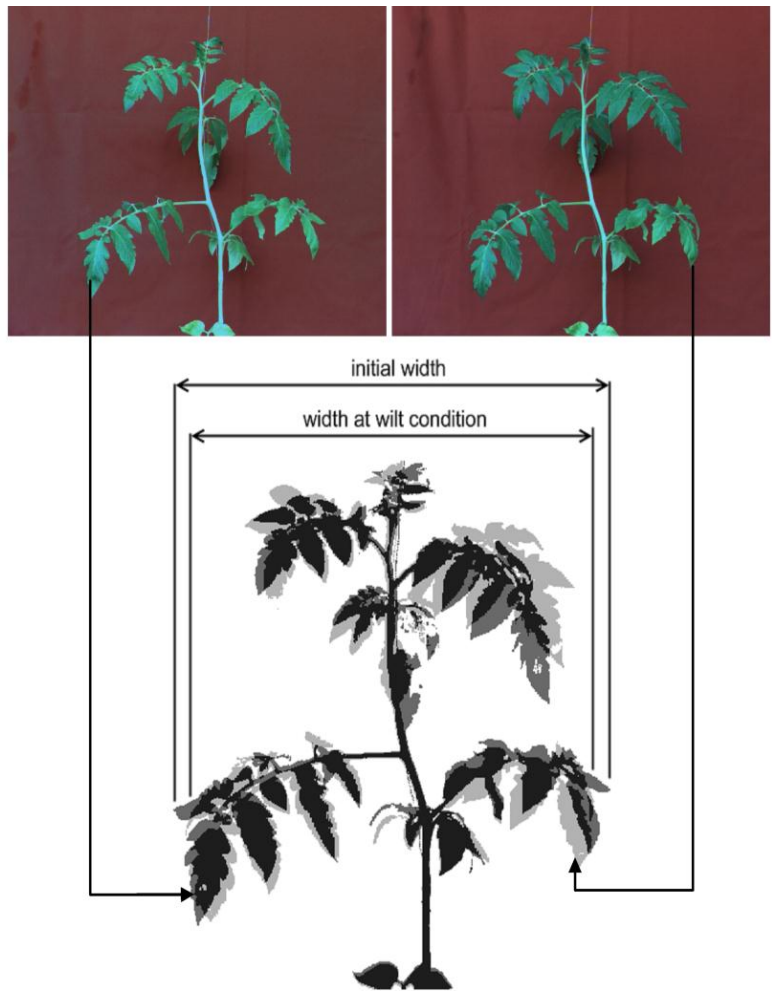

**Figure 4.    Measurement of comparison of width of the plant in image (dark gray is original condition, light gray is wilt condition, and black is overlapped part of the two images)**

For wilt condition determination, ratio value of 2% or more for width comparison found to be reasonable. This means, plant will be watered when its canopy shrinks more than 2% compared to its original canopy in the morning, measured from the width of the plant in the images. Measurement of width difference of the plant from two images is illustrated in Fig. 4.

After determining the ratio value for width comparison, we need to apply a second rule in wilt definition, to avoid system from watering again in the next action since the condition will not change much in a short time, thus the ratio value may be still bigger than 2%. However, if a small change in wilt condition is noticed (or wilt gradient changes its value from positive to negative), it will help to be used as a second rule for wilt condition definition. It was found that the plant will stop shrinking, or at least slow down the process, 10-15 minutes after watering. This response was used as a second rule, so the rules for watering (turn the pump ON for 3 minutes, then OFF again) are:

if (wilt_degree > 0.02 and previous_width > width)

then pump=ON for 3 minutes;

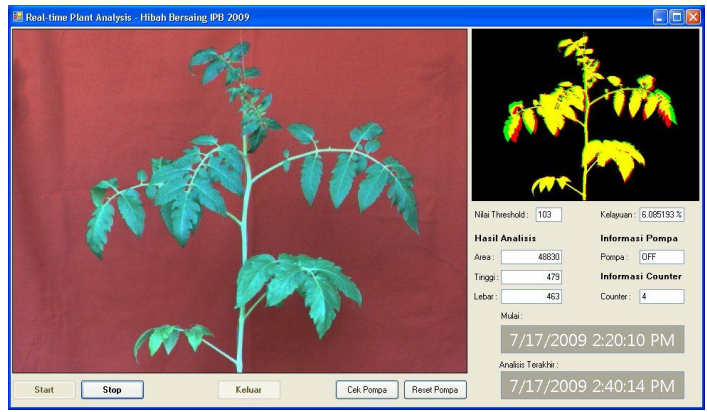

**Figure 5. Developed program for real-time plant monitoring**

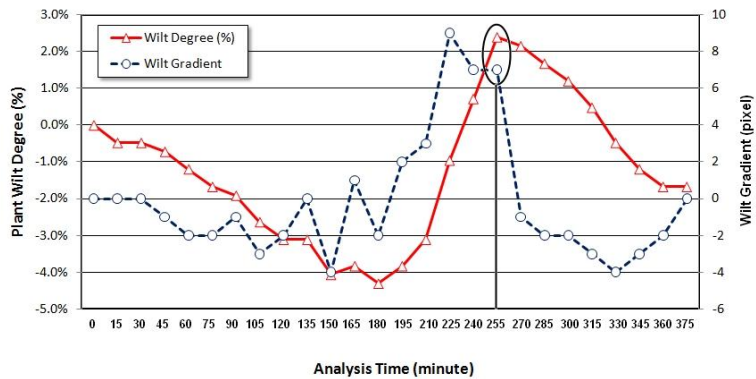

**Figure 6.    The record for 6 hours and 15 minutes real-time monitoring with 15 minutes interval**

The above rules will avoid the automatic fertigation system from watering again in the next action since the condition of the plant will stop wilting and probably get more fresh condition if compared with the last condition just before watering, so that width of plant increasing, not decreasing just like before watering. The image analysis and rules checking were conducted using a developed real-time image processing program (Fig. 5). The program was set up to capture and analysis an image, in 640 by 480 pixels resolution, every 15 minutes. An example of the record for 6 hours and 15 minutes real-time monitoring was shown in Fig. 6. From the figure, it can be noticed that the system only turn on the pump once, after 4 hours and 15 minutes of monitoring (at 255th minute), when the wilt ratio value was more than 2% and the wilt gradient was positive. This time, we only tested that the automatic fertigation system was working and able to flow the water with plant nutrition in it, based on the determined wilt condition which was still very simple. In the next step, determination of wilt condition of the plant being monitored using some other parameters is very important to obtain the real condition of the plant that need watering.

In the application of real-time monitoring system, another 60 plants were grown in the greenhouse and their images were captured every 2 days and analyzed by still image processing to observe their development in height. For the first 30 days, all plants were given nutritive water according to suggested dosage using a timer. The pump was activated every 30 minutes for 3 minutes during daytime. Automatic fertigation based on the developed rules was applied after that for 14 days from 8 am to 5 pm. In the automatic fertigation system, one plant was taken as representative for image capturing and analyzing by the system every 30 minutes, while the action of system was applied to the whole population. At the rest of time until the plants produced fruits, watering was done used the timer again. However, image capturing for height measurement using image processing was conducted until 49 days after planting due to problem with the plants size that caused leaves overlapping so it is very difficult to capture individual plant.

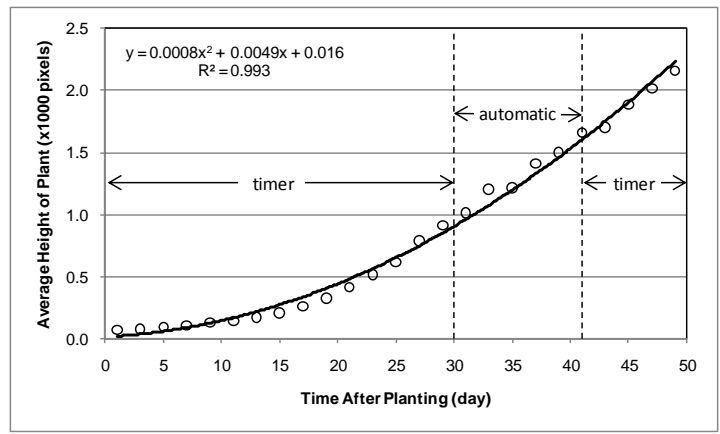

**Figure 7.    Height development of the plants maintained by timer and automatic fertigation**

The results show that automatic fertigation system could maintain the development of the plants normally, the same way with the fertigation applied using the timer (Fig. 7). Advantage of automatic fertigation system application in this experiment is the usage of nutritive water that can be decreased significantly without sacrificing the plants development. For the first 30 days, nutritive water consumed by the population in average was 101.6 liters/day. The consumption of nutritive water drastically decreased to 17.3 liters/day during the 14-days automatic fertigation application, and back again to 101.7 liters/day for the rest of time observed. The different of nutritive water consumption by the population is because when timer was used, nutritive water was given every 30 minutes for 3 minutes (according to suggested dosage), while in automatic fertigation, nutritive water was given only when the plants need them based on plant behavior in response to water shortage. In general, the automatic fertigation system can safe more than 80% of nutritive water if compared with fertigation applied using a timer.

In the future, the research should consider other physical factors from the plant such as color of leaves to provide detail information about micro-nutrient requirement, as well as plant health condition to provide information for any treatment to prevent the plant from being attacked by plant pathogen.

## 5. Conclusions

From this research, some important information has been obtained and can be concluded as follows:

1. Plant growth can be monitored by analyzing a series of images of the plants during their growth period using image processing

2. In the real-time image analysis for automatic fertigation, width of the plant can be used to determine wilt condition by comparing current condition with the original condition when the plant was still fresh.

3. The automatic fertigation system can safe more than 80% of nutritive water compared to timer fertigation.

4. The system should be expanded so that it can provide information about micro-nutrient requirement and information for any treatment to prevent the plant from being attacked by plant pathogen.

## Acknowledgment

The authors would like to thank the Directorate General of Higher Education, Ministry of Education of Indonesia, who funded this research activities through Competitive Grant Program (Hibah Bersaing), 2008-2010.

## Authors

**Usman Ahmad** and **Dewa Made Subrata** are lecturers and researchers in the Dept. of Mechanical and Biosystem Engineering, Faculty of Agricultural Engineering and Technology, Bogor Agricultural University (IPB), Bogor, Indonesia. Usman Ahmad's interest is image processing techniques applications in bioprocess, while Dewa Made Subrata's interest is robotic and automatic control. Some of their works are application of image processing and automatic control in grading machine for agricultural products.

**Chusnul Arif** is a lecturer and researcher in the Dept. of Civil and Environmental Engineering, Faculty of Agricultural Engineering and Technology, Bogor Agricultural University (IPB), Bogor, Indonesia. His research topics are mostly about increasing the efficiency of greenhouse utilization through the employment of new technology.